\documentclass[proof]{WileyASNA-v1}
\UseRawInputEncoding

\articletype{Proceedings}%

\received{}
\revised{}
\accepted{}

\newcommand{\lrp}[1]{\left(#1\right)}
\newcommand{\csi}{\xi}

\begin{document}

\title{Effects of a non-universal IMF and binary parameter correlations on compact binary mergers}

\author[1]{L.M. de S\'{a}}
\author[1]{A. Bernardo}
\author[1]{R.R.A. Bachega}
\author[1,2]{L.S. Rocha}
\author[1]{J.E. Horvath}

\authormark{de S\'{a} \textsc{et al}}

\address[1]{\orgdiv{Universidade de S\~{a}o Paulo}, \orgname{Instituto de Astronomia, Geof\'{i}sica e Ci\^{e}ncias Atmosf\'{e}ricas}, \orgaddress{\city{S\~{a}o Paulo}, \state{SP}, \country{Brazil}}}

\address[2]{\orgname{Max-Planck-Institut f\"{u}r Radioastronomie}, \orgaddress{\city{Bonn}, \country{Germany}}}

\corres{L.M. de S\'{a}\\ \email{lucasmdesa@usp.br}}

\presentaddress{R. do Mat\~{a}o, 1226 - Cidade Universit\'{a}ria, 05508-090, S\~{a}o Paulo-SP, Brazil}

\abstract{Binary population synthesis provides a direct way of studying the effects of different choices of binary evolution models and initial parameter distributions on present-day binary compact merger populations, which can then be compared to empirical properties such as observed merger rates. Samples of zero-age main sequence binaries to be evolved by such codes are typically generated from an universal IMF and simple, uniform, distributions for orbital period $P$, mass ratio $q$ and eccentricity $e$. More recently, however, mounting observational evidence has suggested the non-universality of the IMF and the existence of correlations between binary parameters. In this study, we implement a metallicity- and redshift-dependent IMF alongside correlated distributions for $P$, $q$ and $e$ in order to generate representative populations of binaries at varying redshifts, which are then evolved with the \texttt{COMPAS} code in order to study the variations in merger rates and overall population properties.}


\jnlcitation{\cname{%
\author{L.M. de S\'{a}}, 
\author{A. Bernardo}, 
\author{R.R.A. Bachega}, 
\author{L.S. Rocha}, and 
\author{J.E. Horvath}} (\cyear{2022}), 
\ctitle{Effects of a non-universal IMF and binary parameter correlations on compact binary mergers}, \cjournal{Astron. Nachrichten}, \cvol{}.}

%\fundingInfo{Funding info text.}

\maketitle

%\footnotetext{\textbf{Abbreviations:} ANA, anti-nuclear antibodies; APC, antigen-presenting cells; IRF, interferon regulatory factor}

\section{Introduction}\label{sec1}

The beginning of direct gravitational detections by the LIGO, Virgo and now KAGRA observatories has gradually allowed more constraints to be placed on the properties and origins of binary compact mergers (BCMs). Some of the most useful tools in this regard have been rapid binary population synthesis (BPS) codes, which can perform the isolated evolution of large samples of binary systems from zero-age main sequence (ZAMS) conditions up to a present day population under a wide set of models.

Most often in BPS, the primary mass ($M_1$) is taken to be distributed according to a universal IMF and the mass ratio ($q=M_2/M_1\leq1$), orbital period ($\log P$) and eccentricity ($e$) from uniform distributions or fixed values. However, observational evidence has gradually mounted in support of a more complicated picture \citep{Kroupa2013,Hopkins2018}. The IMF has long been expected to become top-heavy at low metallicities, and suggestions of this behavior have been observed and fitted to an environment-dependent IMF which also relates to the star formation rate (SFR). On the other hand, $q$, $e$ and $P$ have all been shown to be correlated and $M_1$-dependent \citep{Moe2017}. In this work, we have applied both relations in generating ZAMS binary samples to be evolved with the \texttt{COMPAS}\footnote{\url{https://github.com/TeamCOMPAS/COMPAS}} BPS code \citep{COMPAS}, in order to study the cosmic evolution of BCM properties.

\section{Methods}

\subsection{Correlated binary parameters}

\begin{figure*}[ht!]
\centerline{\includegraphics[width=\textwidth]{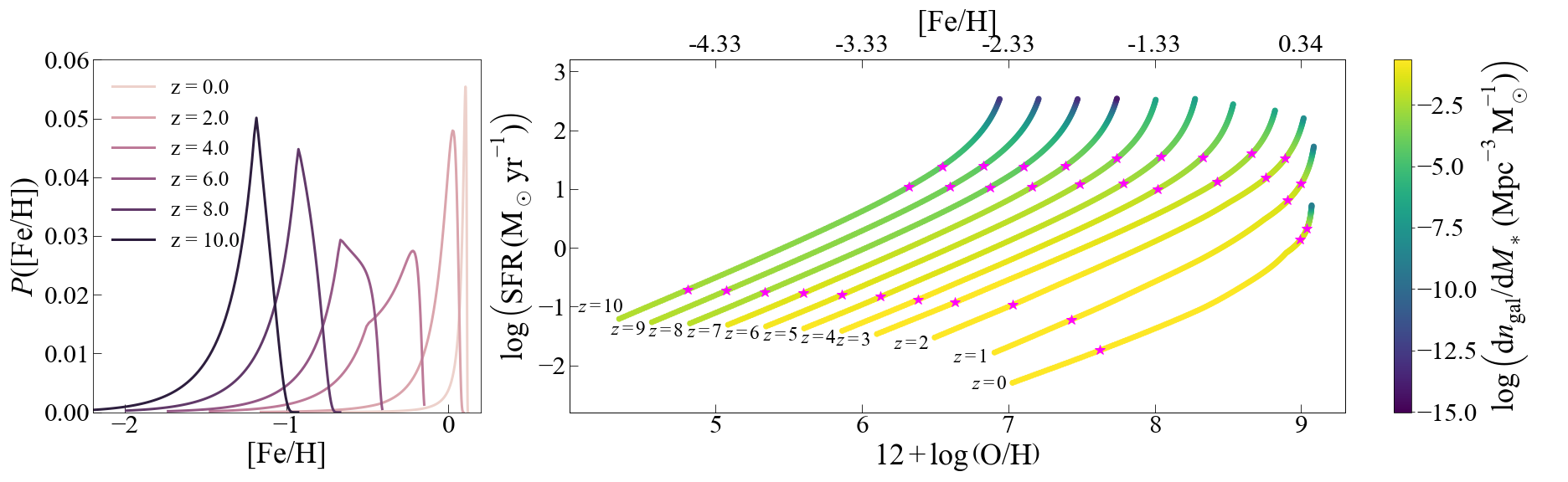}}
\caption{Right: curves representing the resulting metallicity-SFR at different redshifts, colored by the GSMF at each point. Magenta stars show the sampling of three metallicities per redshift according to the GSMF. Left: metallicity distribution per redshift resulting from the MZR, SFMR and GSMF.\label{fig:sfr}}
\end{figure*}

\begin{figure*}[t!]
	\centerline{\includegraphics[width=0.8\linewidth]{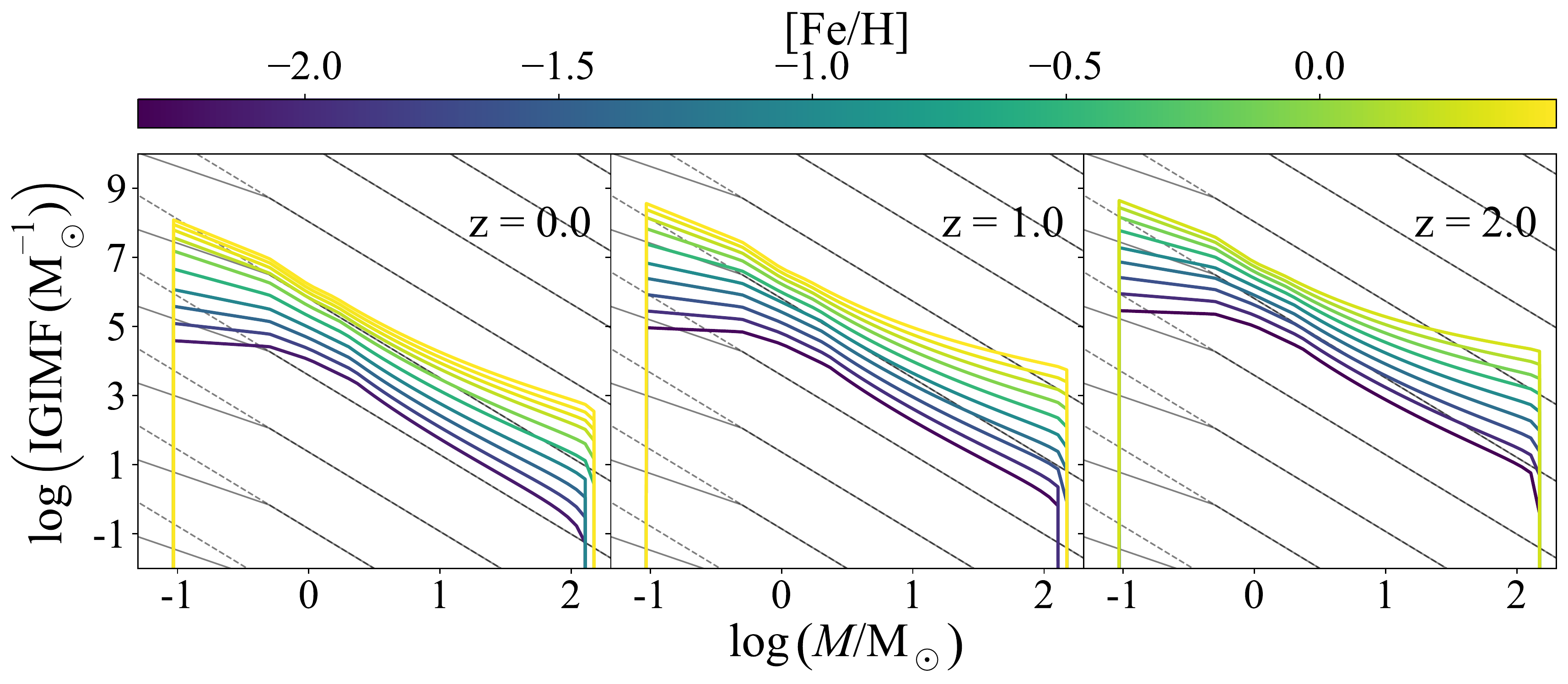}}
	\caption{The varying IGIMF for different metallicities at different redshifts. The dark solid line represents the Kroupa universal IMF, and the dashed line the Salpeter IMF, from which Kroupa deviates at low masses.\label{fig:imf}}
\end{figure*}

For generating our initial samples, we fully adopt the binary parameter distributions from \cite{Moe2017}; due to the distributions' observational limitations, we are restricted to $M_1>0.8\,\mathrm{M}_\odot$, $q>0.1$ and $0.2<\log P<8$ binaries, which should not significantly affect our results for neutron star (NS) and black hole (BH) binaries. For each $M_1$ sampled from the IMF described in the next section, we sample $\log P$ according to their companion frequency, $f_{\log P;q\geq0.1}\lrp{M_1}$, fit. Then, given each $M_1,\log P$ pair, we sample $q$ and $e$ from their respective probability distributions, $p_q\lrp{M_1,\log P}$ and $p_e\lrp{M_1,\log P}$; these are a two-part power law with an excess at $q\geq0.95$ and a simple power-law, respectively. 

\begin{figure*}[t!]
\centerline{\includegraphics[width=0.9\linewidth]{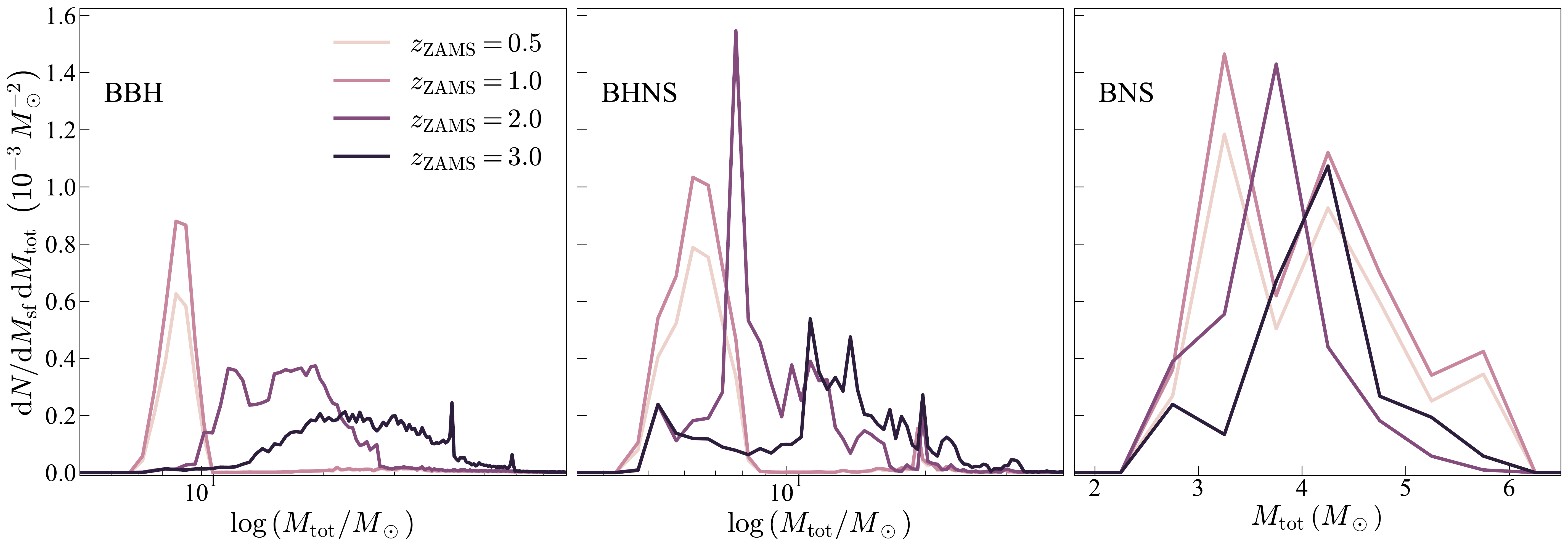}}
\caption{ Total mass distributions for all compact binaries separated by class and $z_\mathrm{ZAMS}$.\label{fig:massdistr}}
\end{figure*}

\subsection{IMF}\label{sec2}

We adopt the integrated galaxy-wide IMF (IGIMF) from \cite{Jerabkova2018}. In the IGIMF approach, stellar formation within a galaxy is considered to take place in embedded clusters (ECLs), individual regions with a given metallicity ($\mathrm{Z}$) and stellar mass ($M_\mathrm{ecl}$) following a specific stellar IMF (sIMF). The \textit{galactic} IMF (gIMF) is then the result of integration over the sIMF of all ECLs. By adopting a composition-dependent sIMF for the ECLs, the gIMF becomes non-universal also.

The metallicity-dependent IMF from \cite{Marks2012}, a three-part power law, is taken as the sIMF ($\csi_\mathrm{s}$) within an ECL; because star formation is constrained by the amount of gas available, the sIMF is a distribution over the mass $M$ of stars of the form $\csi_\mathrm{s}(M|\mathrm{Z}, M_\mathrm{ecl})$, with more massive ECLs producing more massive stars. By considering a galaxy as chemically homogeneous, integration over ECLs becomes an integration over $M_\mathrm{ecl}$. The ECL masses are taken to be distributed according to a power law distribution $\csi_\mathrm{ecl}\lrp{M_\mathrm{ecl}|\mathrm{SFR}}$; it is, like the stellar IMF, limited by the total amount of gas in the galaxy, proportional to the SFR, reflecting the top-heavy IMFs observed at high SFRs by \cite{Gunawardhana2011}. The gIMF is then

\begin{equation}
    \label{eq:IGIFM}
    \csi_\mathrm{g}\lrp{M|\mathrm{SFR},\mathrm{Z}} = \int_0^{\infty}\csi_\mathrm{s}\lrp{M|\text{Z},M_\mathrm{ecl}}\csi_\mathrm{ecl}\lrp{M_\mathrm{ecl}|\mathrm{SFR}}\,\mathrm{d}M_\mathrm{ecl}.
\end{equation}

\subsection{SFR}\label{sec3}

For working with the SFR, we adopt the empirical relations employed by \cite{Chruslinska2019,Chruslinska2020} in calculating their star-formation rate density (SFRD) distribution over metallicity and redshift ($z$). These redshift-dependent relations describe mean galaxy properties as a function of galaxy stellar mass. We adopt their mass-metallicity (MZR) and mass-SFR (SFMR) relations to tie the two quantities at each redshift, while the sampling of (Z, SFR) pairs is made at each $z$ according to the galaxy stellar mass function (GSMF), a galaxy density distribution which links each (Z, z, SFR) to a correspondent star-forming mass ($M_\mathrm{SF}$). We adopt their intermediate-metallicity fiducial model. 

\begin{figure}[t]
	\centerline{\includegraphics[width=\linewidth]{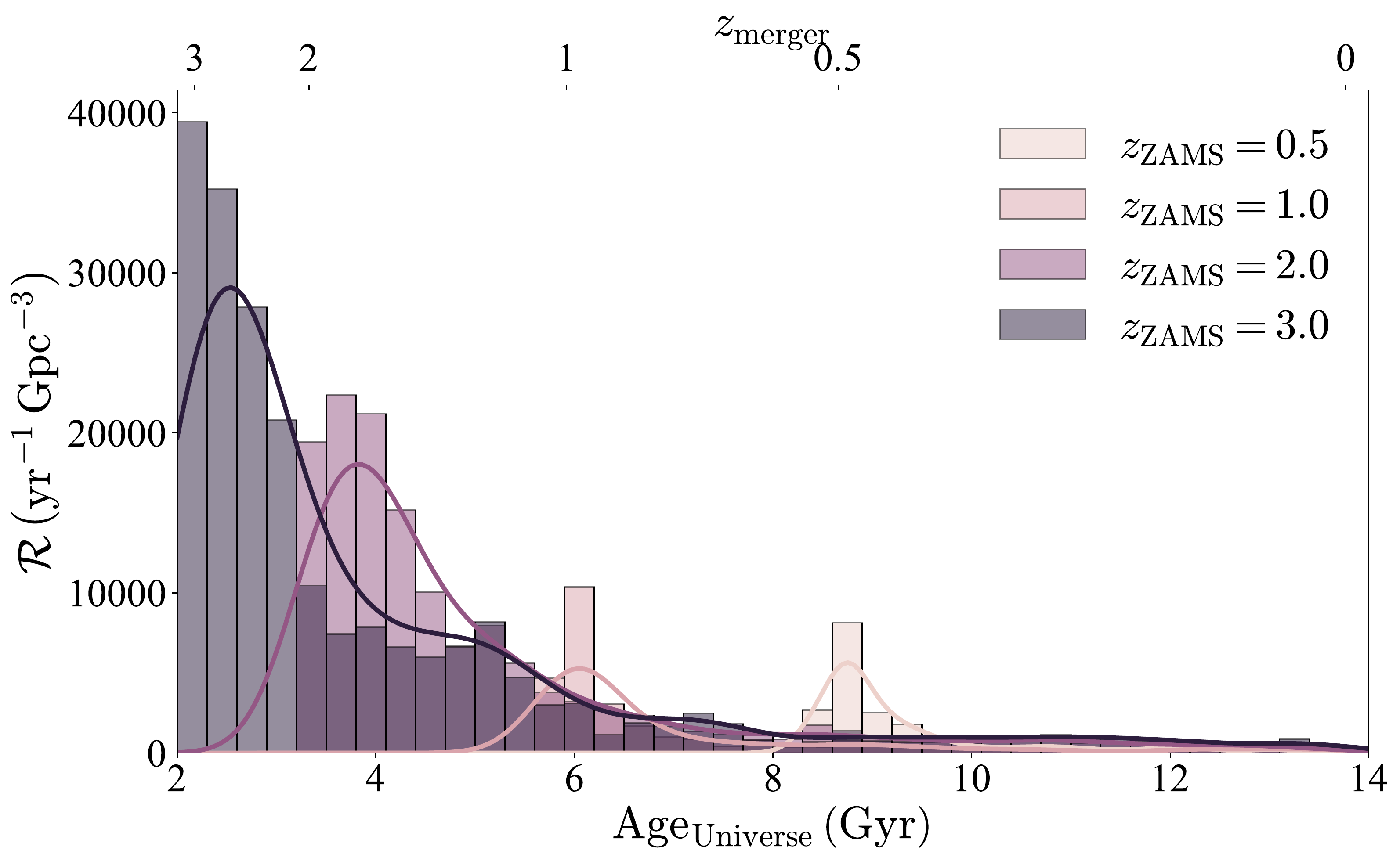}}
	\caption{Merger rate distribution for all CBMs over universe age at merger, for each $z_\mathrm{ZAMS}$.\label{fig:mergerdistr}}
\end{figure}

By matching the SFR, Z and $z$ in this way, we effectively treat the gIMF as $\csi_\mathrm{g}\lrp{M|\mathrm{Z},z}$, allowing us to build samples representative of conditions observed at any given $z=z_\mathrm{ZAMS}$.

\section{Results}\label{sec4}

The results presented here have been derived from 12 samples of $\approx10^6$ binaries each, for three metallicities sampled by stellar mass density at each $z_\mathrm{ZAMS}=0.5,1,2,3$; they are indicated by the magenta stars at the corresponding redshifts in Figure \ref{fig:sfr}. These were then evolved for $13.8\,\mathrm{Gyr}$ in \texttt{COMPAS}. In the following we introduce some preliminary results.

In Figure \ref{fig:massdistr}, we show the total mass ($M_\mathrm{tot}$) distribution for all compact binaries by type and $z_\mathrm{ZAMS}$, as the binary count per $M_\mathrm{tot}$ bin, per $M_\mathrm{sf}$. While all classes show a tendency for more massive systems to have been formed earlier, this is particularly strong for BHs. Although it is possible that the NS progenitor mass range is not as strongly redshift-dependent as that of BH progenitors, we cannot yet discard the possibility that this is a product of our BNS sample being currently about $10$ times smaller than the BBH and BHNS samples.

\begin{figure}[t!]
	\centerline{\includegraphics[width=\linewidth]{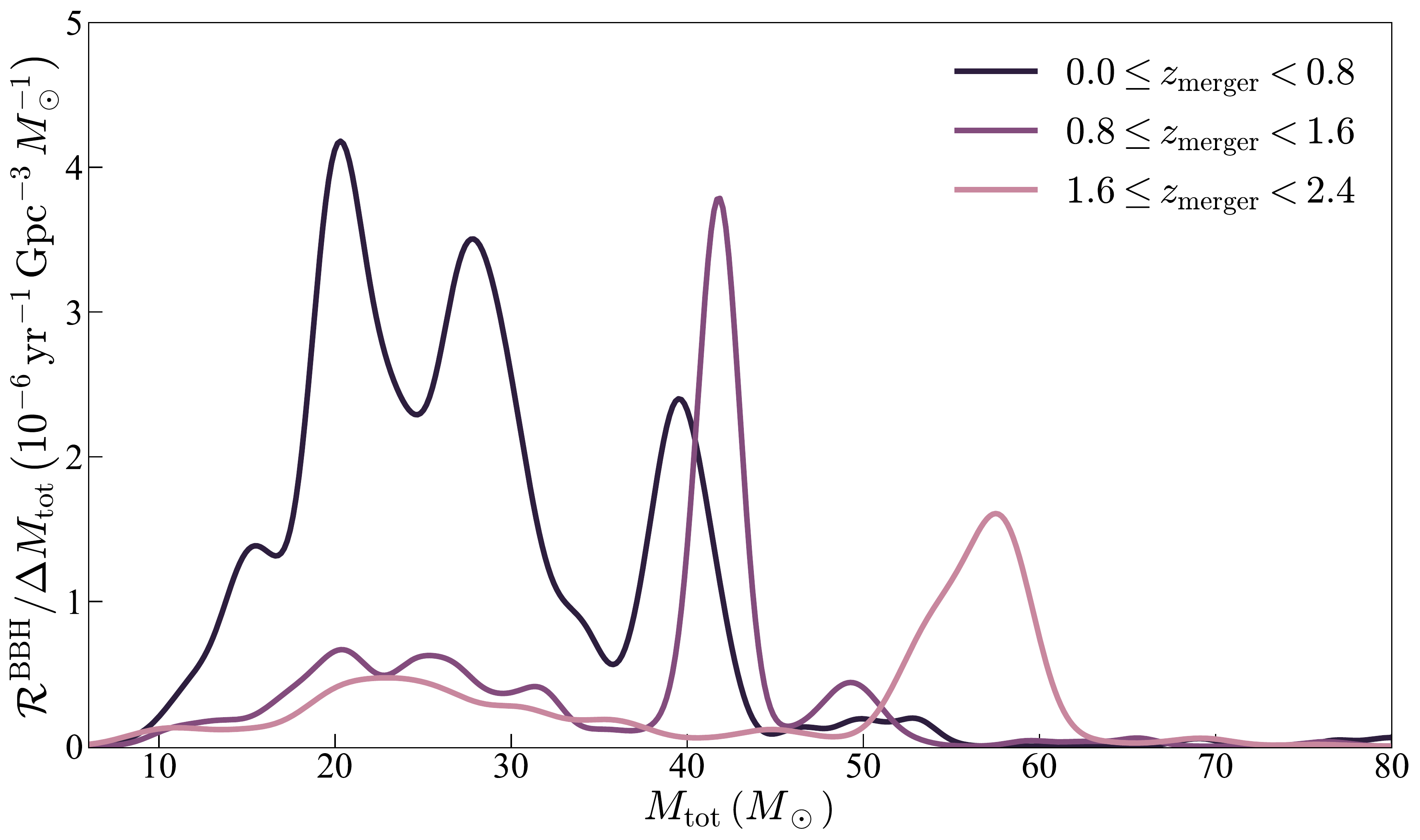}}
	\caption{BBH merger rate per total mass bin distribution, for three different $z_\mathrm{merger}$ ranges.\label{fig:bhbhrate}}
\end{figure}

In Figure \ref{fig:mergerdistr} we show the merger rate per comoving volume distribution over redshift/Universe age at merger, for all BCMs, for each $z_\mathrm{ZAMS}$. While younger systems naturally merge later, we also see a strong increase in the peak merger rate with $z_\mathrm{ZAMS}$ and in the overall merger rate with $z_\mathrm{merger}$. This indicates that a much greater merger rate in the past is related to a tendency of older populations to produce more BCMs, which could be linked to a top-heavy IMF at large redshift.

We show also the overall binary black hole (BBH) merger rate, $\mathcal{R}^\mathrm{BBH}$, distribution with $z_\mathrm{merger}$ in Figure \ref{fig:bhbhrate}, where a strong preference for massive mergers can be seen with increasing $z_\mathrm{merger}$. This behavior suggests that, if observed, such a mass-redshift dependence could be linked to a non-universal IMF. On the other hand, support for the gIMF employed here indicates that the total mass of BCMs could be applied as a good distance proxy.

Finally, we report in Table \ref{table} our merger rates at $z_\mathrm{merger}=0$. We note that only $\mathcal{R}^\mathrm{BBH}$ falls outside of the current LIGO 90\% credibility interval. While this in principle could indicate a preference for a different metallicity distribution over our fiducial model, it is also possible that this result is related to two effects not yet fully incorporated into the synthesis.

The first is the matter of sample resolution. Because the merger rate varies with redshift and metallicity, its distribution over them is only approximated by a discrete set of configurations. With our current grid, it is possible that the approximated distribution has been smoothed out in such a way as to overestimate the merger rate. The second effect is the formation of higher-order multiples which should be taken into account in the total star-forming mass $M_\mathrm{SF}$. Our computation of $M_\mathrm{SF}$, however, considers that every star that is not in a binary is isolated, missing the masses of companions in triples, quadruples and so on. Once this is accounted for, we can expect a decrease in the resulting volumetric rates.

\begin{center}
\begin{tabular}{cccc}
\toprule
Source &    \multicolumn{3}{c}{Local merger rate $(\mathrm{yr}^{-1}\,\mathrm{Gpc}^{-3})$} \\
{} & BBH & BHNS & BNS \\
\midrule
         This work & $432$ & $124$ & $63$  \\
         GWTC-3 & $16-130$ & $7.4-320$ & $13-1900$  \\
\bottomrule
\end{tabular}
\captionof{table}{Local merger rates from this work and 90 \% credibility intervals from GWTC-3.}
\label{table}
\end{center}

\section{Conclusions}\label{sec5}

Is this work we have sought to fully incorporate both a non-universal, cosmically evolving, gIMF and empirically-fitted correlated binary parameter distributions into the sampling of ZAMS binaries to be evolved with the \texttt{COMPAS} BPS code. By sampling environmental conditions according to the GSMF at different redshifts, we have been able distinguish populations by age and compute time-evolving merger rates. We have been able to estimate from it merger rates of $432$, $124$ and $63\text{ yr}^{-1}\text{ Gpc}^{-3}$ for BBH, BHNS and BNS mergers, the last two of which are compatible with the most recent LIGO estimates. We have also found that an IMF that becomes top-heavy at large redshifts leads to a strong preference of more massive BCMs for higher redshifts. A full exploration of compact binary parameters and evaluation of the effects of \cite{Moe2017}'s parameter correlations remains to be performed.

%\backmatter

\section*{Acknowledgments}

Financial support was provided by the \fundingAgency{Funda\c{c}\~{a}o de Amparo \`{a} Pesquisa do Estado de S\~{a}o Paulo (FAPESP)} through the grants \fundingNumber{13/26258-4} and \fundingNumber{2020/08518-2}. LMS acknowledges the \fundingAgency{Conselho Nacional de Desenvolvimento Cient\'{i}fico e Tecnol\'{o}gico (CNPq)} for financial support, grant number \fundingNumber{140794/2021-2}.

\nocite{*}% Show all bib entries - both cited and uncited; comment this line to view only cited bib entries;
\bibliography{Wiley-ASNA}%

\end{document}